\documentstyle[floats,preprint,aps,epsf]{revtex}

\draft
 
\newcommand{\be}{\begin{equation}}
\newcommand{\ee}{\end{equation}}
\newcommand{\ba}{\begin{eqnarray}}
\newcommand{\ea}{\end{eqnarray}}

\begin{document}

\title{ Chaos and Taub-NUT related spacetimes}

\author{ P.S. Letelier\footnote{e-mail: letelier@ime.unicamp.br}
 and 
W. M. Vieira\footnote{e-mail: vieira@ime.unicamp.br} } 
 
\address{
 Departamento de Matem\'atica Aplicada-IMECC,
Universidade Estadual de Campinas,
13081-970 Campinas,  S.P., Brazil}
\maketitle
 
\begin{abstract}
 
 The occurrence of chaos for  test particles 
moving  in a  Taub-NUT  spacetime with 
a dipolar halo perturbation is studied using Poincar\'e sections. We find 
 that the NUT parameter (magnetic mass)  attenuates the presence of chaos.
 \end{abstract}

The Taub-NUT (Newman, Tamburino  and Unti) spacetime \cite{taub,nut} is 
one of the  most  bizarre solutions of the vacuum Einstein
 equations. Because of its
  many unusual properties it has been
described as a ``counterexample to almost anything" \cite{misner}. It has
 closed timelike curves, is nonsingular in a meaningful mathematical 
sense but is not geodesically complete, etc. For certain range  of the
 coordinates, it can
 be seen as a  Schwarzschild monopole endowed with 
a ``magnetic mass" \cite{ash}. 

The Euclidean version of this metric  has recently received 
some attention due to the fact that   it is closely related to the dynamics 
of two non-relativistic Bogomol'nyi-Prasad-Sommerfield (BPS)
 monopoles \cite{gary}. The
asymptotic motion of monopoles corresponds to geodesic motion in
 Euclidean Taub-NUT space, this motion is integrable. This fact 
has motivated the study of geodesics in  Euclidean Taub-NUT
 and related spaces \cite{vis}.

Examples of chaotic motion
in General Relativity are the geodesic motion of a test particle moving in the geometry associated with : a)   Fixed two body
  problem  \cite{cont}, b)  A monopolar 
center of attraction surrounded by a dipolar halo \cite{wldi,rotbh} (in
 Newtonian theory this system is  integrable),  c)  A monopolar 
center of attraction surrounded by a  quadrupole plus octupole
  halo \cite{wlqo}, d)  Multi-Curzon and multi-Zipoi-Vorhees  solutions \cite{ssmae}, and
   e) A rotating black hole (Kerr geometry) with a dipolar halo \cite{rotbh}.
Also  gravitational waves can produce irregular motion of test particles
 orbiting around a static black hole \cite{bc,lwmelgr}. 
 
In this Letter we consider the geodesic motion of test particles
 moving in  a Taub-NUT spacetime perturbed by a distant  distribution
 of matter that can be
represented by a dipole. The case of a center of attraction
 (without NUT parameter) perturbed by a dipolar halo  was studied
 in \cite{wldi}, the combined relativistic effects and the
 breakdown of the reflection symmetry in this
 case produces a
non integrable motion.  Our main goal in this letter is to study 
 the effect that the magnetic mass
has on the  chaotic motion of test particles.
The geodesic motion in a pure Taub-NUT spacetime was studied 
in \cite{mistaub}, this case is completely integrable.
Also generalizations and perturbations of this spacetime  has
 been considered from the view point of spacetime dynamics \cite{monc}.

The metric that represents the superposition of a  Taub-NUT metric and
  a dipole along the $z$-axis is a stationary axially symmetric
 spacetime. The vacuum Einstein equations for this class of
  spacetimes is an 
integrable system of equations 
that is closely related to the principal sigma model
 \cite{cosgrove}. Techniques to actually find the solutions
 are B\"acklund 
transformations and the inverse scattering method, also 
 a third method  constructed with elements of the previous
 two is the ``vesture method", all these methods are 
closely related \cite{cosgrove}.
 The general metric that represents the nonlinear superposition of
a Kerr-NUT   solution with a Weyl solution, in particular,
 with a multipolar expansion can be found by  using the ``inverse scattering 
  method"  \cite{sol}. For the particular case of a Taub-NUT
 metric with a dipolar halo  we find 
\be 
ds^2=g_{tt}(r,z)dt^2+2 g_{t\phi}(r,z)dt d\phi+g_{\phi\phi}(r,z)d\phi^2+
 f(r,z)(dz^2+dr^2),
\label{m1}
\ee  
where

\ba
&&g_{tt}=-  e^{-2 {\em D} u v } [ e^{-4 {\em D} u}(e^{-8 {\em D} v 
}(m+1)^{2}(u^2-1)+e^{-8 {\em D}} (m -1)^{2}(u^2-1)\nonumber\\
&& \hspace{1cm}+2 e^{-4 (v  +1){\em D}}(u^2-v^2 )b^{2}
  )  +e^{-4(v +1 ){\em D}}(e^{-8 {\em D} u} +1)(v^2-1) b^{2}]/H\nonumber\\
&&g_{t\varphi}=-[2 e^{-2 {\em D}(u+v+1)}(e^{-4 {\em D}u}(e^{-4 {\em D}v}(m+1)
(u-v)(u +1)\nonumber\\
&& \hspace{1cm}-e^{-4 {\em D}}(m-1)(u+v)(u -1))(v +1)+e^{-4 {\em
 D}v}(m +1)(u+v )\nonumber\\
&&\hspace{1cm}(u +1)(v-1)-e^{-4 {\em D}}(m-1)(u-v)(u-1)(v -1)) b
 ]/H \nonumber\\ 
&&g_{\varphi\varphi}=
- \sigma^2(e^{-2{\em D}(2u-uv)}(e^{-8 {\em D} v}(m+1)^{2}(u+1)^{4}
 +e^{-8 {\em D}}(m-1)^{2}  (u-1)^{4}\nonumber\\
&& \hspace{1cm}+2e^{-4 {\em D}( v+1)}(u^2-v^2)( u^2-1) b^{2}
)((v^2-1)\nonumber\\
 && \hspace{1cm}       +e^{-4 {\em D}( v+1)} (e^{-8 {\em D}
 u}(v+1)^{4}+(v-1)^{4})(u^2-1)b^{2})/H  \nonumber\\
&&f= \frac{\sigma^2}{4}H\exp[{\em D}^{2}(u^2 v^2- u^{2}- v^{2}
 +1)+2 {\em D}( u v+2u+2v+2)] \nonumber\\
&&H\equiv e^{-4 {\em D} u}[e^{-8{\em D}v}(m+1)^{2}(u +1)^{2}+e^{-8 
{\em D}}(m-1)^{2}(u -1)^{2}\nonumber\\
&& \hspace{1cm} +2 e^{-4{\em D}(v+1)}(u^2-v^2)b^{2}]+e^{-4 {\em
 D}(2 u+ v +1}  (v +1)^{2} b^{2}+e^{-4 {\em D}( v-1)}(v-1)^{2} b^{2} \nonumber\\
&& \sigma^2\equiv m^2+b^2 \label{BZ}
 \ea

 The coordinates $(r,\phi, z)$ are dimensionless and have the range of
 the usual cylindrical  coordinates. They are related to $u$ 
and $v$ by: $z=uv$ and
 $r=(u^2-1)^{1/2}(1-v^2)^{1/2}$, $u\geq 1$ and $-1\leq v \leq -1$. Our
units are such that  $c=G=1$; $m$   {\it D} and $b$  represent the mass, 
 the halo dipole
 strength,  and $b$ the NUT parameter, respectively. The constant $\sigma$
 represents the  geometric ``sum" of the ``electric" and ``magnetic" mass.
 The coordinate transformation $ t'=t, \, u=R/m-1, \, v=\cos\vartheta, \, \varphi^\prime=\varphi$
 reduces (\ref{BZ}) with $D=0$ to the  Taub-NUT solution in the usual spherical 
 coordinates \cite{BZ}.

To study the small dipole perturbation case is better to  use the 
metric  obtained by keeping the   first 
order terms in  the dipolar strength  
 {\em D} in the exact metric (\ref{BZ}).  This approximation, for
 the parameters and
 range of coordinates used, will not produce a
 significant information loss; we shall comeback to this point later. We
 find for $g_{\mu\nu}=g_{\mu\nu}^0 +{\em D}  g_{\mu\nu}^1 $,
 \ba
&&g_{tt}=-(2(4 m^{2} v -2 m^{2} +2 m u v+u^{2}v-3v +2){\em D} u-F
         )(1-u^2 ))/F^{2} \nonumber\\
&&g_{t\varphi}=2 [-2( (  (3 v^{2}-1)u^{2}+(2v-3)(v+1)u^{4}-v^{2}+v) m 
-2(u^2-v^2)m^{2}u\nonumber\\
&&\hspace{1cm}+(5 v^{2} -2 v-1 ) u^{3} -(3 v-2 )u v -u^{5} )
 {\em D}-F(u^2 -1) v]b/F^2 \nonumber\\
&&g_{\varphi\varphi}=- \sigma^2[2 (2 (2 ((2 v^{2}-v+1)-(v +1) u^{2} )m^{2} u+ (u^{4}
 v+3 u^{4}+6 u^{2}v -2 u^{2}\nonumber\\
&&\hspace{1cm}+v-1)(v-1 )m-(5 v-1) u +(v -1 ) u^{5}+4 u^{3} v^{2})
          F(v+1)-(2 ((2 v+1) u^{2} +v-1) m\nonumber\\
&&\hspace{1cm}+(v+2)u^{3}-2 m^{2}u
            +5 u v)(4 (u^{2} +1)(v^2-1 ) m u-4 (u^2-v^2) m^{2}+2 
            (3 v^{2}-1)u^{2}\nonumber\\
&&\hspace{1cm}+(v^2 -1)u^{4}-3 v^{2}-1))(-1){\em D}+(4 (u^{2} +1)
             (v^2-1) m u-4 (u^2-v^2)  m^{2}\nonumber\\
&&\hspace{1cm}+2 (3 v^{2} -1) u^{2} +
          (v^2 -1 )u^{4}-3 v^{2}-1 )F]/F^2 \nonumber\\
&&f =\sigma^2(2 (2  (u^{2} -v +1 ) m - (3 v -2) u+2 m^{2} u +u^{3} 
v) {\em D} +F) \nonumber\\
&&F\equiv 1+2 m u +u^{2}  \label{LT}
\ea
The geodesic equations for the metric (\ref{m1}) can be cast as
\ba
&&\dot{t}=g^{tb}E_b,\hspace{0.5cm} \dot{\phi}=g^{\phi b}E_b , \label{geo1} \\
&&\ddot{r}=-\frac{1}{2f}[g^{ab}_{,r}E_a E_b +f_{,r}(\dot{r}^2
-\dot{z}^2)+2f_{,z}\dot{r}\dot{z}], \label{geo2}\\
&&\ddot{z}=-\frac{1}{2f}[g^{ab}_{,z}E_a E_b +f_{,z}(\dot{z}^2
-\dot{r}^2)+2f_{,r}\dot{r}\dot{z}], \label{geo3}
\ea 
where the dots denote derivation with respect to $s$ and the indices
$a$ and $b$  take the values $(t,\phi)$, $g^{ab}$ stands for 
 the inverse of $g_{ab}$. $E_t=-E$ and $E_\phi=L$ are integration 
constants; $E$ and $L$ are the test  particle energy and angular momentum,
 respectively.
The set (\ref{geo1})--(\ref{geo3}) admits a third integration constant
\be
E_3=g^{ab}E_a E_b+f(\dot{r}^2+\dot{z}^2)=-1 . \label{e3}
\ee %
Thus to have complete integrability we need one more independent constant
 of integration. 
In the case of pure Taub-NUT solution (${\it D}=0$) we have a fourth
constant due to the existence of a third Killing vector and  a fifth
(Runge-Lenz vector) related to the existence of a Killing-Yano tensor.

 The  system   (\ref{geo2})--(\ref{geo3}) can be written as a four
 dimensional dynamical system in the variables $(r, z, P_r= 
\dot{r}, P_z=\dot{z})$. A  convenient method to study 
qualitative aspects of this system
  is to compute the Poincar\'e sections through  the plane
  $z=0$. The intersection  of the orbits with this plane will be
 studied in some  detail for bounded motions. We shall
 numerically solve the system  (\ref{geo1})--(\ref{geo3}) and use 
the integral (\ref{e3})
to control the accumulated error along the integration; we
 shall return to this point later.

 The Poincar\'e section for different
initial conditions    with 
  energy $E=0.967$ and angular momentum $L=3.75$ 
moving in an  exact Taub-NUT 
geometry (${\it D}=0$) with NUT parameter $b=0.28$ and ``total mass" $\sigma=1$
 are presented  in Fig. 1.   We have the typical section of an integrable motion, i.e.,  the sectioning   of invariant tori, for integrability and KAM
 theory, see for instance \cite{arnIII}. The values  of $\sigma, E$ and $L$ will be kept unchanged in our numerical analysis.

The motion of test particles around 
 a Kerr and a Schwarzschild black hole with a 
 dipolar halo ($b=0$ and ${\it D}\not=0$ ) is
 chaotic and it was studied in some detail in  \cite{rotbh,wldi} for
 different energy shells.  In Fig. 2 we show
the Poincar\'e section for ${\it D}=0.0005$ and the same values of
  $E=0.967 $ and $L=3.75$ as in Fig. 1. We find  islands of integrability 
surrounded  by chaotic motion. The two isolated islands
around the points (10, 0.05) and (5, 0.075) correspond to  the same
 torus. In the case studied in \cite{wldi}  they were closer.

 Now we shall consider  a  particle moving around  a Taub-NUT
attractive 
center surrounded by a weak  dipolar halo. In Fig. 3 we draw the Poincar\'e section for   
${\it D}=0.0005$, $b=0.14107$, $\sigma=1$,
$E=0.967$ and $L=3.75$ (direct rotation). We see that the islands
of stability are larger in this case than in perturbed Schwarzschild solution 
 (see Fig. 2);
also we have new systems of small islands immersed in the 
chaotic region.  We
 have that the chaotic region is smaller
in this case than in the  perturbed Taub-NUT case. It does 
look like that the presence of the NUT parameter diminishes the effect
 of the dipolar  strength as a chaos source. In Fig. 4 we present
 the  section with the same parameters
 of Fig. 3, except that now we have increased the NUT
parameter, $b=0.254787$. In  Fig. 4 we observe that  the integrable region  increases in such a way that
the chaotic region has almost disappear  in this scale. In other words,  the
 NUT parameter has the property of restoring the invariant tori. In 
Fig. 5 we present a magnification of the region around the
 point $(0.0744,7.37)$ of Fig. 4, we find a small chaotic region
 around the  crossing of the section around the indicated point.
Although in the present Letter we present results 
for particular values of the parameters involved, we did a rather 
extended  numerical study that supports  our conclusions, Figs. 3, 4, and 5
  being representative of this search. 
The change of sign of the angular particle angular momentum does not
 alter the
figures and the change of sign of ${\it D}$ introduces only  a $P_r \rightarrow
-P_r$ transformation.

 For the values of the parameters  ${\it D}=0.0005 $, $b=0.14107, 0.254787$
$E=0.967$ and $L=\pm 3.75$, we have that the particles move roughly 
in the ``box"
$4.7<r<21$, $-6< z <7$. We take as a  measure of error the quantities 
\be
\Delta g_{ab}=|(g_{ab}^{ex}-g_{ab})/g_{ab}^{ex}|,  \hspace{0.5cm}
\Delta f= |(f^{ex}-f)/f^{ex}|, \label{del}
\ee 
where  in  these expressions the   sum rule of repeated 
indices does not apply.   $g_{\mu\nu}^{ex}$  and $ g_{\mu\nu}$
 refer to the solutions (\ref{BZ}) and (\ref{LT}),  respectively.
We find that for the above mentioned range of coordinates 
and the values of parameters used in this Letter the quantities 
defined in (\ref{del}) are   at most of  the order
of $10^{-4}$. Also in this range,  the error in the derivatives of 
the metric functions  is even smaller (the metric functions are
 very smooth).  We also want to mention that  the Poincar\'e sections 
shown in this Letter were computed
 with  an accumulated error in the ``energy" [cf. Eq. (\ref{e3})]
 smaller than  $10^{-10}$.

In summary the NUT parameter has the rather surprising property of making the 
motion of particles  more integrable, it enlarges the region of invariant tori.
This is a puzzling result since the addition of the NUT parameter  and a dipole moment to a black hole makes the metric rather complicate, hence one should 
foresee more complex motion with a greater  destruction of tori. It happens the opposite. This is again a manifestation  of Misner characterization of the Taub-NUT metric as a `counterexample to almost anything".

The authors thank CNPq and FAPESP for support and to S.R. Oliveira   for discussions.

\newpage
\noindent
{\bf FIGURE CAPTIONS}

\vspace*{1cm}
\noindent
Fig. 1.    Poincar\'e section of  test 
particles moving with angular momentum $L=3.75$ and energy $E=0.967$ in a
 Taub-NUT geometry with NUT parameter $b=0.28$  and 
 $\sigma=1$.  This is a typical section of an integrable system. \\
 
\vspace*{0.4cm}
\noindent
Fig. 2.  Poincar\'e section for ${\it D}=0.0005$,
  $E=0.967$, $L= 3.75$, $b=0$, and  $\sigma=m=1$.  The two isolated
 islands
around the points (10, 0.05) and (5, 0.075) are parts of the same
 torus.\\ 

\vspace*{0.4cm}
\noindent
Fig. 3.  Poincar\'e section for   ${\it D}=0.0005$, $b=0.14107$, $\sigma=1$
$E=0.967$ and $L=3.75.$ The islands
of stability are larger in this case than in the geometry without NUT parameter
  (cf. Fig. 2).\\

\vspace*{0.4cm}
\noindent
Fig. 4. Poincar\'e section with the same parameters
 of Fig. 3, except that now $b=0.254787$. The chaotic 
region  is almost  inexistent.\\

\vspace*{0.4cm}
\noindent
Fig. 5. The region around the point $(0.0744,7.37)$ of Fig. 4 is magnified. We find a small chaotic region.

\noindent

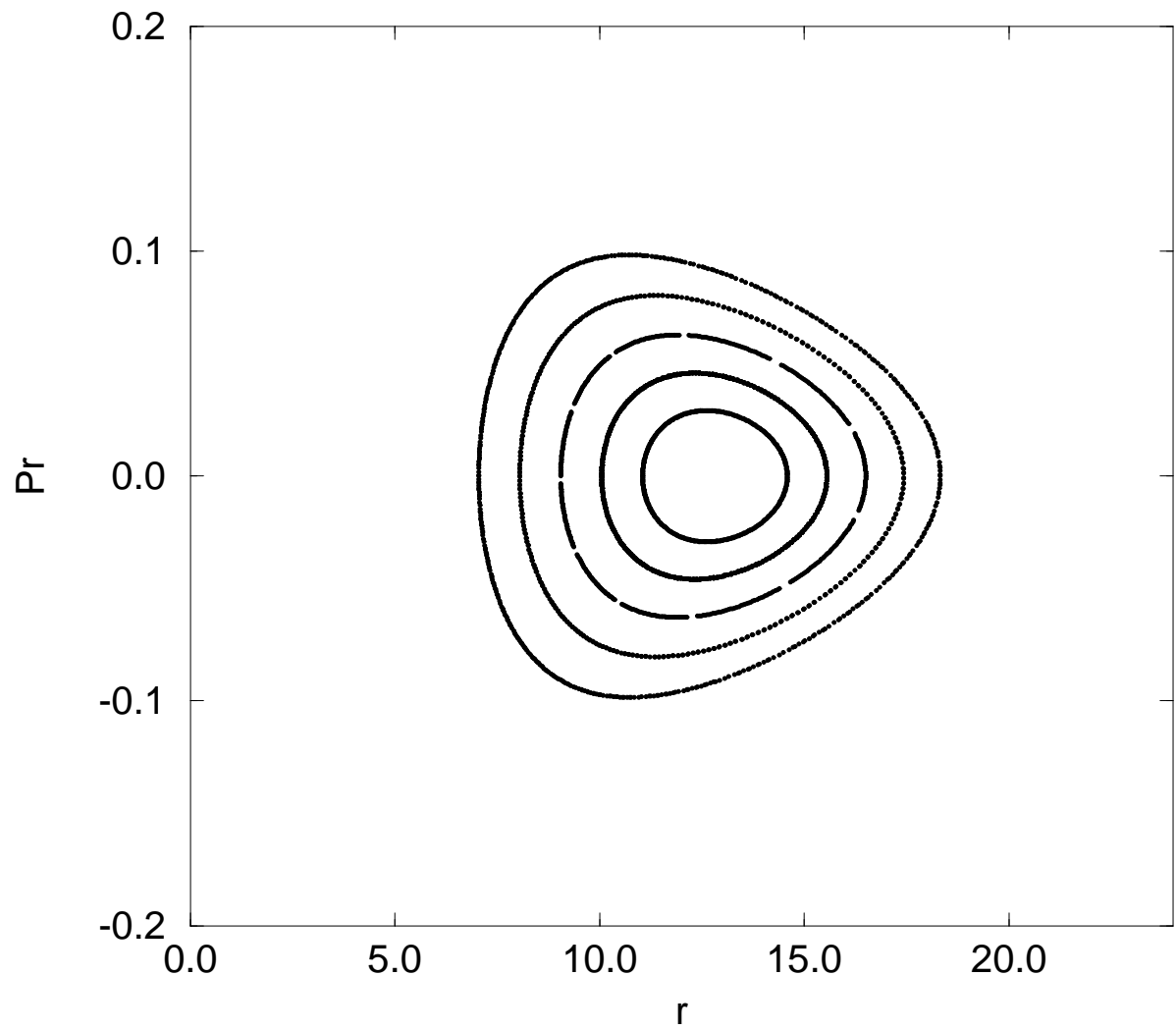
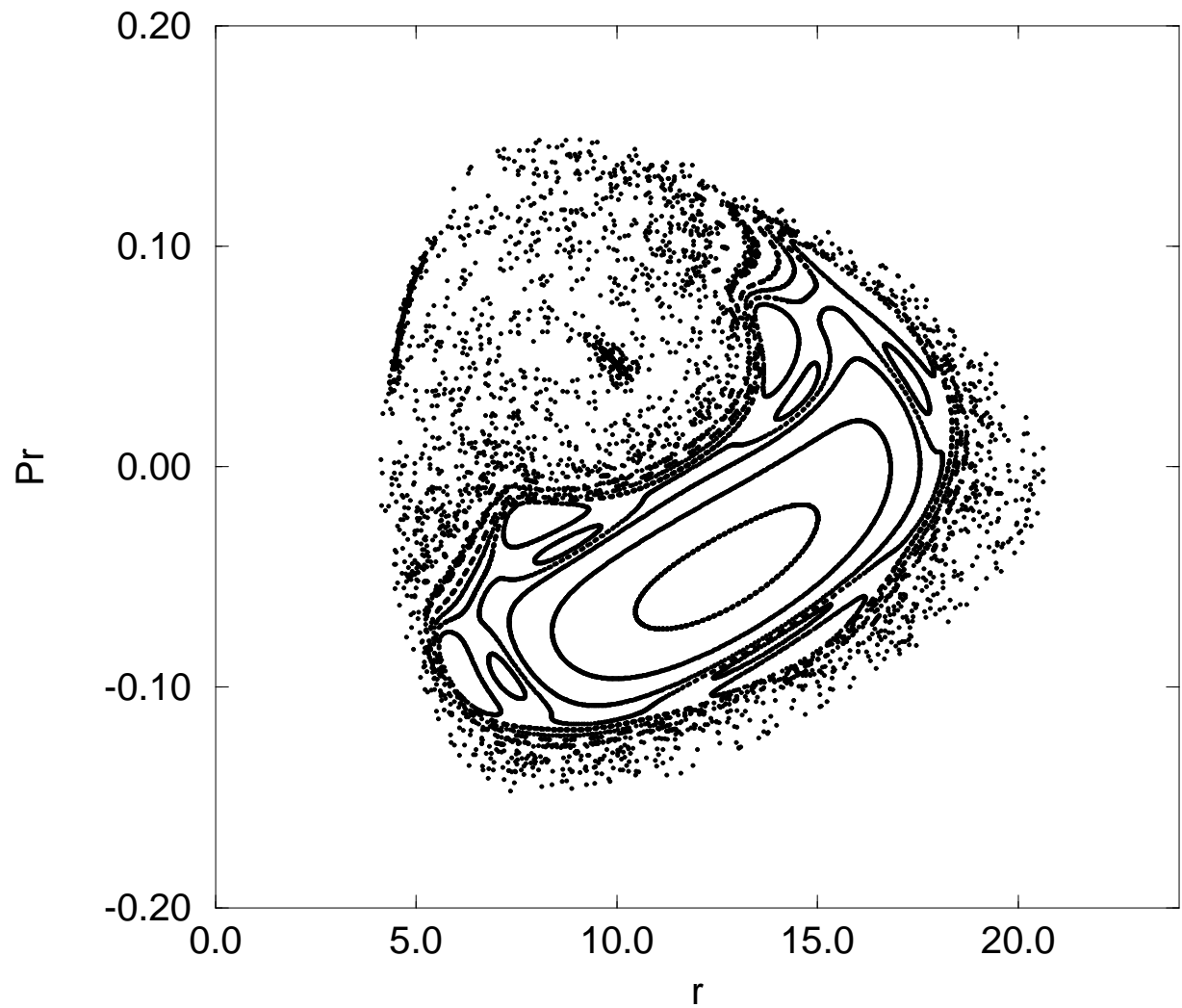
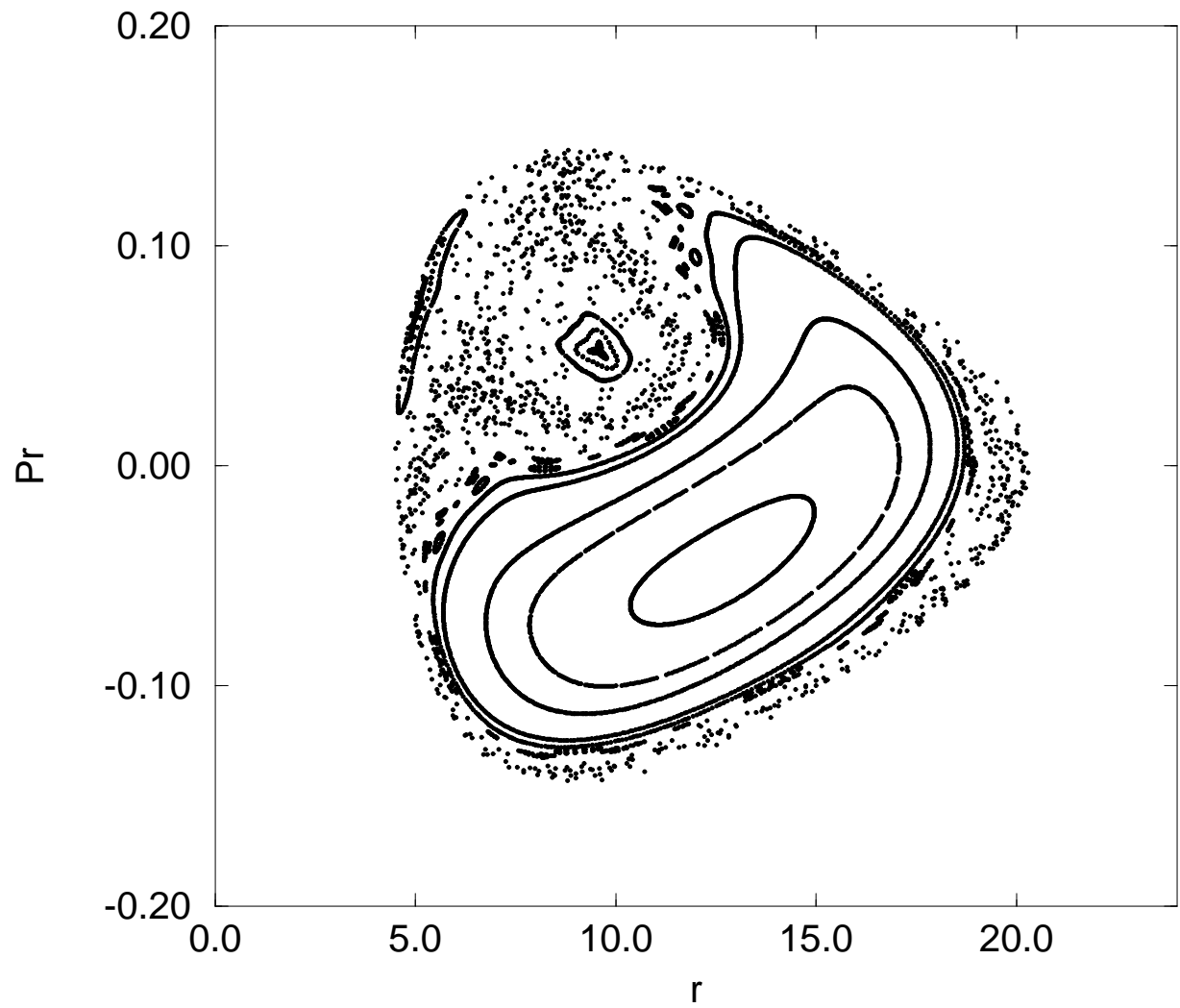
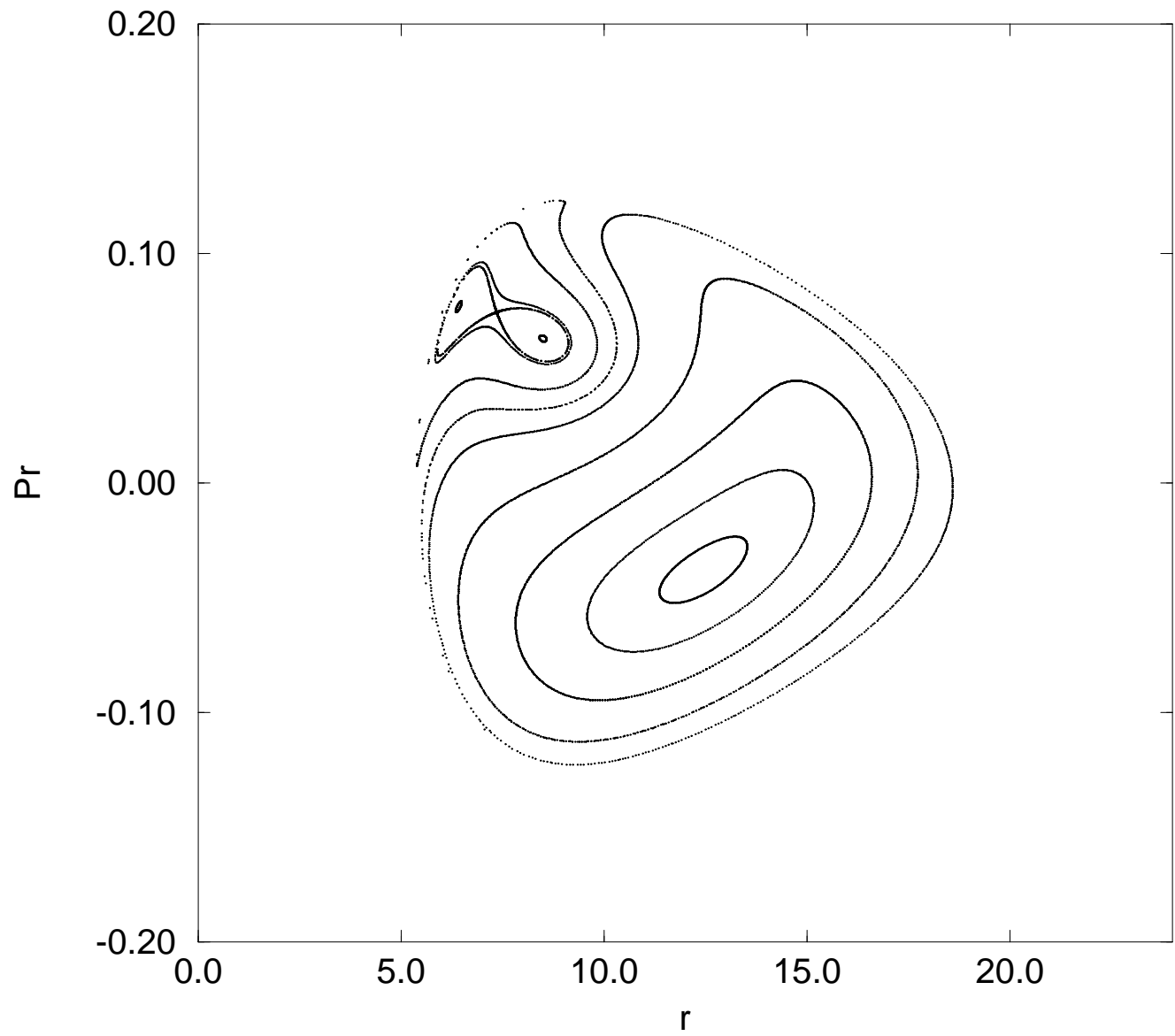
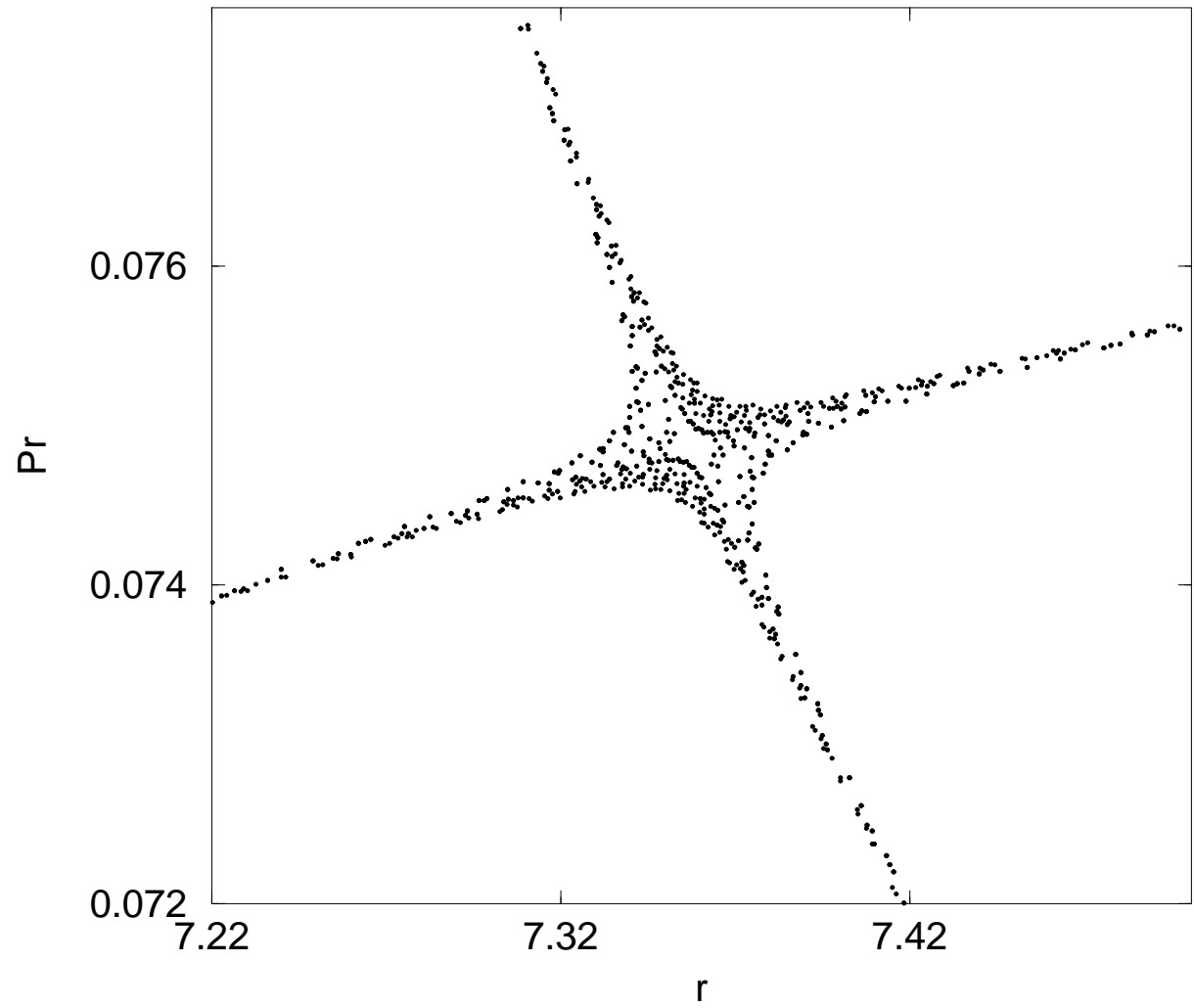
\end{document}